\documentclass[aps,prb,10pt,twocolumn,superscriptaddress,amsmath,amssymb,showpacs,reprint]{revtex4-2}
\usepackage{textcomp}
\usepackage{graphicx}
\usepackage{dcolumn}
\usepackage[absolute,overlay]{textpos}

\usepackage{bm}
\usepackage[colorlinks, allcolors=blue]{hyperref}
\usepackage[usenames, dvipsnames]{color}
\usepackage{setspace}
\usepackage{amssymb}
\usepackage{scalerel}
\usepackage[utf8]{inputenc}
\usepackage[T1]{fontenc}
\usepackage{hyperref}
\usepackage{natbib}
\usepackage[normalem]{ulem}
\usepackage{cancel}
\usepackage{amsmath}
\usepackage{amsfonts}
\usepackage{stackengine}
\usepackage{comment}
\usepackage[scr=esstix,cal=boondox]{mathalfa}
\usepackage{mathrsfs}
\usepackage{csquotes}
\usepackage{soul}
\usepackage[normalem]{ulem}
\usepackage{xcolor}
\usepackage{placeins}

\stackMath
\newsavebox\tmpbox

\begin{document}
\preprint{APS/123-QED}

\title{Thermal robustness of sensing quantum phases of matter via qubit probes}
\author{Sambunath Das}
\email{sambunath@fzu.cz}
\affiliation{FZU - Institute of Physics of the Czech Academy of Sciences, Na Slovance 1999/2, 182 00 Prague, Czech Republic}
\author{Hari Kumar Yadalam}
\email{hari\_kumar.yadalam@kcl.ac.uk}
\affiliation{Department of Physics, King’s College London, Strand, London, WC2R 2LS, United Kingdom}
\affiliation{School of Physics, Trinity College Dublin, Dublin 2, Ireland}
\author{Mark T. Mitchison}
\email{mark.mitchison@kcl.ac.uk}
\affiliation{Department of Physics, King’s College London, Strand, London, WC2R 2LS, United Kingdom}
\author{Katarzyna Roszak}
\email{roszak@fzu.cz}
\affiliation{FZU - Institute of Physics of the Czech Academy of Sciences, Na Slovance 1999/2, 182 00 Prague, Czech Republic}
\date{\today}

\begin{abstract}
We study the effect of temperature on the ability of a qubit probe to distinguish between different phases in strongly correlated systems. Now to this end, we investigate a spin-1/2 Heisenberg XXZ chain coupled to a qubit probe via a pure dephasing interaction. At zero temperature, as reported previously, the decoherence dynamics exhibit different behavior in the gapped and gapless phases of the chain, with characteristic strong oscillations in the gapped phase and monotonic decay in the gapless phase. These oscillations decay much more slowly than expected with rising temperature in the gapped phase and their remnants are still visible at infinite temperature. In the gapless phase at low temperatures, using Luttinger liquid theory, we predict that the coherence exhibits asymptotic exponential decay in time, with the decay rate sensitive to the sign of the anisotropy parameter. We conclude that probe dynamics continue to carry information about the chain even at finite, reasonably small temperatures, positioning qubit probes as sensitive and robust detectors for quantum phases of correlated spin systems.
\end{abstract}

\maketitle

\section{\label{sec:intro}Introduction}
Decoherence is typically considered an obstacle for quantum information processing and computing tasks~\cite{David_1999, Zurek_2003, Schlosshauer_2005,preskill2018quantum,Zurek_2025}. 
However, the decoherence of a small system (probe) can encode valuable information about a larger system which is treated as its environment and is hard to access experimentally. This area of research is significant for both 
theoretical and practical applications. 
The decoherence dynamics of quantum probes have been extensively utilized for thermometry~\cite{Johnson_2016, Razavian_2019, mehboudi2019thermometry,Mitchison_2020, widera2022thermometry, Sindre_2024}. It has also been used for quantum sensing~\cite{Degen_2017,Rovny_2024,lukin2011metrology,bayat2025metrology}, and the study of quantum phases~\cite{Mark_2016, Elliott_2016,Chatterjee_2019} and quantum phase transitions (QPTs)~\cite{Marcin_2025}.

The quantum phases of matter in strongly correlated systems are influenced by many-body interactions~\cite{Subir_2011}, reduced dimensionality~\cite{Giamarchi_2003}, frustration from competing interactions~\cite{Lacroix_2011}, strong coupling between spin and charge degrees of freedom~\cite{Imada_1998} and external tuning parameters such as pressure, temperature and magnetic field~\cite{Imada_1998,Gegenwart_2008}. In low dimensional spin chains and ladders, geometrical confinement enhances quantum fluctuations, while strong correlations suppress classical order~\cite{Giamarchi_2003}. This interplay
of many factors gives rise to a variety of quantum exotic phases, such as the gapped and gapless phases~\cite{Haldane_1983}, spin liquid phase~\cite{Broholm_2020,Savary_2017,Patrik_2008}, vector chiral phase~\cite{Chubukov_1991,Hikihara_2008,Aslam_2016,Das_PRB_2021,Das_JAP_2021}, dimer order phase~\cite{Chubukov_1991} and topologically nontrivial ground states ~\cite{Levin_2006,Hannes_2019,Vijay_2015}. 

Identifying different phases using traditional experimental techniques remains difficult, as they typically assess bulk-averaged features, which may obscure local quantum behavior~\cite{Slichter_1990, Ernst_1987, Reif_2020, Zavoisky_1944, progress_2021, Jaros_2024}. They are also insensitive to quantum coherence and entanglement and generally lack the spatial and energy resolution required to investigate low-energy excitations and short-range correlations. In this context, methods involving qubit spin probes offer a promising approach, enabling local, high-resolution measurements of nonequilibrium quantum dynamics~\cite{Hossein_2026} and thermodynamics~\cite{Mazzola_2013, Dorner_2013}, correlations~\cite{ Elliott_2016,Rodriguez_2018} and quantum phases~\cite{Mark_2016,Chatterjee_2019} and the associated QPTs~\cite{Benedetti_2014, Adam_2022, Franchini_2017, Machado_2023,  Nicolas_2023, Zhang_2008,Marcin_2025}. These QPTs occur at zero temperature and are not driven by thermal fluctuations, but rather by quantum fluctuations~\cite{Giamarchi_2003,Subir_2011}. 

However, precise measurements at zero temperature in experiments are prohibited by the third law of thermodynamics~\cite{Zemansky_1981,Callen_1985}, hence QPTs cannot be directly accessed at zero temperature. Consequently, a significant  effort has been directed towards understanding whether and what survives of such transitions at finite temperature~\cite{Werlang_2010, Karpat_2014,Li_2020,Nandi_2022}. Such remnant signatures have been probed in experiments with a range of techniques; particularly relevant to this work is the technique that employs nitrogen-vacancy (NV) centers in diamond as qubits to probe many-body systems in their proximity~\cite{Ju_2014,You_2014,Patel_2024, Rovny_2024}. Most theoretical works analyzing such scenarios assume zero temperature~\cite{Marcin_2025, Quan_2006, Cucchietti_2007, Rossini_2007}. Notably, there exists a study investigating QPT signatures in the dipolar spin-ring model accessed via a qubit probe at non-zero temperatures, but it focuses on second-order QPTs and considers a qubit probe that is globally coupled to the chain~\cite{Chen_2013}. Given the atomic resolution of NV center probes and the importance of QPTs which are not second order, here we explore how raising the temperature of a many-body system affects qubit decoherence. We further assess whether, despite strong thermal fluctuations at high temperatures, the qubit probe can still reveal the system’s quantum phases and phase transitions. 

Toward this goal, we consider the spin-1/2 XXZ nearest-neighbor Heisenberg chain~\cite{Bonner_1964,Yang_1966_Jul, Yang_1966_Oct,Woynarovich_1987} at thermal equilibrium as a paradigmatic many-body system serving as the environment. The
Heisenberg model is exactly solvable via the Bethe ansatz~\cite{Bethe_1931, Mattis_1993} and is important for studying quantum magnetism and integrability in low-dimensional systems~\cite{Giamarchi_2003,Subir_2011,Franchini_2017}. The model is also simple and concrete enough to allow us to perform detailed numerical analysis, but rich in the thermodynamic limit, showing first-order and infinite-order phase transitions in the ground state, separating two gapped phases by a gapless phase~\cite{Giamarchi_2003,Franchini_2017}. 

At zero temperature, the entire spin chain is in its ground state. Coupling a single qubit to the middle site of the chain, 
via an interaction that leads to pure dephasing of the qubit~\cite{zurek03_i,roszak20i,strzalka21},
encodes information about the chain into the dynamics of the probe's coherences. This enables probing qualitative chain features such as quantum phases, phase transitions, and information propagation~\cite{Marcin_2025}. Specifically, qubit decoherence is qualitatively different in the gapped and gapless phases. In the gapless phases we observe a decay of coherence characteristic for an interaction  with a large environment, while in the gapped phase oscillations are observed because the qubit's environment effectively behaves like a two-level system. Furthermore, the transition between the two types of decoherence is gradual when the anisotropy parameter is negative and abrupt when it is positive, so the change in decoherence reflects the way that the energy gap opens in the energy spectrum of the chain.

In this work, we study qubit decoherence at finite temperatures and observe that the qubit dynamics not only modify the coherence decay rates but also change the functional form of the decoherence, while still exhibiting distinct behavior in both gapped and gapless phases. As expected, the oscillations characteristic of the gapped phase are damped with temperature,
since more states of the chain are involved in the qubit-chain evolution at finite temperatures. However, some distinction between the two types of decoherence can be made even
at infinite temperatures since the oscillatory behavior persists. This shows that the chain effectively acts as a sparse environment in the gapped phase even at high temperatures. Using Luttinger liquid theory, we predict distinct temperature dependences the of the decoherence rate in the attractive and repulsive regimes of the gapless phase. It increases linearly with temperature in the attractive regime, while following a nonlinear power law dependence in the repulsive regime. Thus we show that the qubit probe can also be used to detect different quantum phases of the chain at realistic temperatures, making the method a good candidate 
for experimental application. 

This paper is organized as follows. In Section~\ref{sec:model}, we introduce the model Hamiltonian describing the dynamics of our system and briefly review the measure of qubit decoherence. In Section~\ref{sec:results}, we present the results of finite-temperature calculations for both the gapped and gapless phases corresponding to negative values of the anisotropy parameter, followed by a discussion of the results. We conclude the paper in Section~\ref{sec:summary}. Additional technical details and supplementary derivations are provided in Appendices~\ref{sec:Appendix1}--\ref{sec:Appendix4}. We work in units with $\hbar=1$ and $k_B=1$ throughout.

\begin{figure}
	\begin{center}
		\includegraphics[width=3.4 in]{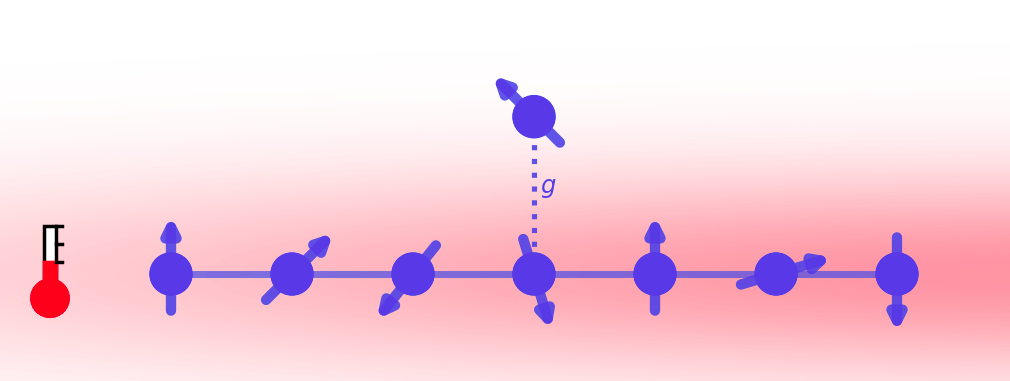}
		\caption{Schematic diagram of a single qubit-probe interacting with the central site of a thermal spin chain. Here, $g$ is the probe-chain coupling strength.}
		\label{fig1}
	\end{center}
\end{figure}

\section{\label{sec:model}Model and Methods}

We study the evolution of the probe qubit interacting with the central site of an antiferromagnetic Heisenberg XXZ spin
chain as depicted in Fig.~\ref{fig1}.
The total Hamiltonian associated with this composite system is given by 
\begin{equation}
\label{total}
       H = H_{C}+ H_{Q}+ H_{CQ},
\end{equation}
where the first term is the Hamiltonian of the spin chain, the second term is the qubit Hamiltonian, and the third term is the interaction between the qubit and the spin chain.

The Hamiltonian of the spin-1/2 Heisenberg chain with open boundary conditions is given 
by~\cite{Giamarchi_2003}
\begin{equation}
\label{xxz}
	H_\mathrm C=J\sum_{k=1}^{L-1}\left[\frac{1}{2}\left(S^{+}_{k} S^{-}_{k+1}+S^{-}_{k} S^{+}_{k+1}\right)+\Delta S^z_{k} S^z_{k+1}\right],
\end{equation}
where the exchange interaction $J$ along the chain is considered positive, $\Delta$ is the anisotropic parameter, $\mathbf{S}_{k}$ is the spin-$1/2$ operator of the $k$-th spin, and $L$ is the length of the spin chain. The sign of $\Delta$ determines whether interactions between spin excitations are attractive ($\Delta <0$) or repulsive ($\Delta >0$), while for $\Delta =0$ the system is effectively noninteracting. The system exhibits three different zero-temperature phases in the thermodynamic limit: for $\Delta <-1$ the ground state is ferromagnetic and gapped, for $-1 < \Delta \leq 1$ it is a gapless Luttinger liquid, whereas for $\Delta>1$ it is antiferromagnetic and gapped.

The qubit-chain interaction is given by
\begin{equation}
\label{int}
	H_{CQ} = \frac{g}{2}\sigma_z S_{M}^z,
\end{equation}
where $M=L/2$ denotes the middle spin of the chain, the coupling constant is fixed at $g=1/4 J$ following Ref.~\cite{Marcin_2025},
and the chain energy parameter $J$ is treated as a constant and used to define all units. 
The free qubit Hamiltonian is given by
\begin{equation}
H_\mathrm{Q}=\epsilon\,\frac{\sigma_z}{2}, 
\end{equation}
where $\sigma_z$ is the appropriate Pauli matrix and $\epsilon$ is the energy splitting of the probe qubit.  
Since the interaction and the qubit Hamiltonians commute (are diagonal in the same basis $\{|0\rangle,|1\rangle\}$)~\cite{roszak18i}, the interaction can only lead to qubit pure dephasing, so
only qubit coherence (the off-diagonal elements of the density matrix
written in the $\{|0\rangle,|1\rangle\}$ basis) evolves due to the presence of the chain~\cite{Roszak_2015}. 

The qubit is initially prepared in an equal superposition of states $|0\rangle$ and $|1\rangle$,
\begin{equation}
\psi_{Q}= \frac{1}{\sqrt{2}}(|0\rangle+|1\rangle),
\end{equation}
to maximize the coherence.
The initial state of the environment is given by the Gibbs state,
\begin{equation}
\label{ini}
\rho_{C}(0)=e^{-\beta H_{C}}/\mathrm{Tr}\left[e^{-\beta H_{C}}\right].
\end{equation}
The introduction of temperature ($T = 1/\beta$) in the case of the Heisenberg spin-1/2 XXZ chain leads to additional dynamical features, enriching the qubit’s evolution and rendering the model more physically realistic. The temperature determines the initial state of the spin chain: instead of being in a pure ground state, the chain is prepared in a thermal (mixed) state described by the Gibbs ensemble. This thermal initialization introduces statistical fluctuations in the chain degrees of freedom, which in turn, influence the qubit’s decoherence dynamics. 

The evolution of the qubit coherence is given by~\cite{Thomas_2006}
\begin{equation}
\label{eq:lecho}
\rho_{01}(t)=\langle 0 |\rho_{Q}(t)|1\rangle=\frac{e^{i\epsilon t}}{2}
\langle e^{i (H_{C}+ \frac{g}{2}S_{M}^z) t}e^{-i (H_{C}- \frac{g}{2}S_{M}^z)t} \rangle,
\end{equation}
where the expectation value is taken over the initial state of the chain (\ref{ini}).
We note that for $\epsilon=0$, due to spin-inversion symmetry of the thermal state and the spin-chain Hamiltonian, $\rho_{01}(t)$ remains real throughout the evolution (see appendix~\ref{sec:Appendix1} for proof). In the following the free qubit energy 
is always set to $\epsilon=0$. This is because the evolution only introduces trivial 
oscillations that have no bearing on the physics under study.

In the following, we numerically compute $\rho_{01}(t)$ using Eq.~\eqref{eq:lecho} for spin-chain of length $L=64$ for different values of spin-chain anisotropy parameter ($\Delta$) and its temperature ($T$). This is done by purifying the mixed state of the spin chain by introducing an identical copy (ancilla), and evolving the matrix product state~\cite{Schollwock_2011,Orus_2014} approximation of this purified state using the time-dependent variational principle algorithm~\cite{Kramer_2008,Haegeman_2011,Haegeman_2016,Mingru_2020}. A more detailed description of the purification method is given in the appendix~\ref{sec:Appendix2}. In the next section, we discuss the results obtained using this methodology. Numerical convergence checks of these results are given in the appendix~\ref{sec:Appendix3}.

\section{\label{sec:results}Results}

\begin{figure*}[!ht]
	\begin{center}
		\includegraphics[width=6.2 in]{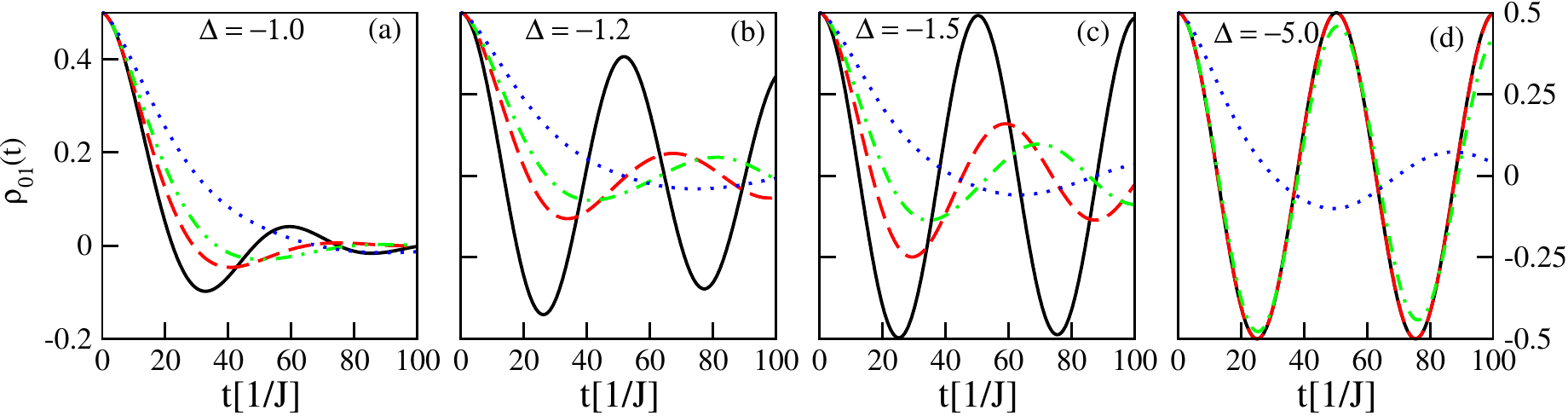}
		\caption{Qubit coherence $\rho_{01}(t)$ for a chain of length $L = 64$ and bond dimension $\chi = 768$ in the ferromagnetic gapped phase ($\Delta < -1$) with (a) $\Delta=-1$, (b) $\Delta=-1.2$, (c) $\Delta=-1.5$, (d) $\Delta=-5.0$. Different curves correspond to different temperatures: $T/J = 0.1$ (solid black lines), $T/J = 0.5$ (dashed red lines), $T/J = 1.0$ (dotted-dashed green lines), and $T/J = \infty$ (dotted blue lines). Note that (b) and (c) are plotted using the same scales as (a), whereas (d) is plotted using a different vertical scale.}
		\label{fig3}
	\end{center}
\end{figure*}
\begin{figure*}[!ht]
	\begin{center}
		\includegraphics[width=6.2 in]{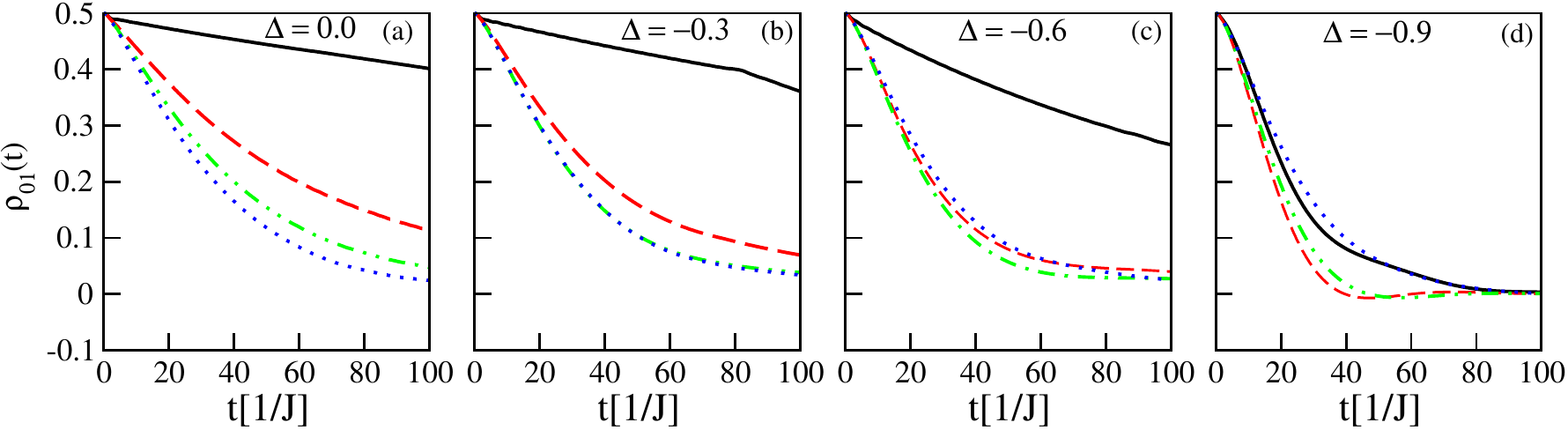}
		\caption{
			Qubit coherence $\rho_{01}(t)$ for a chain of length $L = 64$ and bond dimension $\chi = 768$ in the gapless phase with non-positive anisotropy parameter ($-1<\Delta\le 0$) with (a) $\Delta=0$, (b) $\Delta=-0.3$, (c) $\Delta=-0.6$, (d) $\Delta=-0.9$. Different curves correspond to different temperatures: $T/J = 0.1$ (solid black lines), $T/J = 0.5$ (dashed red lines), $T/J = 1.0$ (dotted-dashed green lines), and $T/J = \infty$ (dotted blue lines). Figures (b), (c) and (d) are plotted using the same scales as figure (a).}
		\label{fig4}
	\end{center}
\end{figure*}

In this section, we present and discuss results showing the effect of thermal fluctuations of the spin-chain on the qubit probe dynamics in both the non interacting ($\Delta = 0$) and attractive ($\Delta < 0$) regimes of the spin chain. The results in the repulsive case ($\Delta > 0$) are presented in the appendix~\ref{sec:Appendix4}.

For the attractive interactions, $\Delta<0$, it is energetically favorable for the z-components of spins on neighboring sites to remain aligned parallel to each other, just like in the classical Ising model. Yet the hopping term in the Hamiltonian introduces quantum fluctuations between the spin configurations.  At zero temperature ($T=0$), these competing effects give rise to a first order QPT at $\Delta=-1$, separating the symmetry-broken ferromagnetic gapped phase ($\Delta<-1$) and the symmetric gapless Luttinger liquid phase ($-1<\Delta < 0$)~\cite{Subir_2011,Giamarchi_2003}. These two-phases have a qualitatively distinct excitation spectrum above the ground state: magnon excitations in the gapped phase; and spinon excitations in the gapless phase. In previous work by some of us~\cite{Marcin_2025}, we showed that the qualitative change of the spin-chain state across this zero temperature QPT can be probed through the dynamics of a qubit probe coupled to it. We found that the qubit coherence exhibits qualitatively distinct behavior in the two phases separated by this QPT: qubit coherence shows oscillatory behavior in the gapped phase and monotonic decay in the gapless phase.

However, the Hohenberg-Mermin-Wagner theorem~\cite{Halperin_2019} forbids such transitions at finite temperatures in 1D quantum systems with short-range interactions and continuous symmetries (like the spin-chain model considered), so the zero-temperature transition becomes a smooth crossover. Since the excitation spectrum is distinct in both the phases, the question arises if it is still possible to distinguish between the two phases by observing the decoherence dynamics of the qubit at non-zero temperatures. Below we show that it is indeed possible to distinguish this, since the distinct oscillation characteristic for the gapped phase persists even at high temperatures. The amplitude of the oscillations is damped by temperature, but the qualitative difference between the two types of evolutions remains pronounced at relatively high temperatures, contrary to naive expectation.

In order to understand the temperature dependence of the qubit dynamics in the ferromagnetic gapped phase ($\Delta<-1$), the temperature should be compared to the excitation gap of the chain, which sets the relevant energy scale. The many-body spectral gap of the spin-chain is given by  $E_{gap}= |\Delta|-1 $~\cite{Giamarchi_2003}. The ratio of the temperature to this gap determines the finite temperature behaviour. In this work, we define the low-temperature regime as $T\ll E_{gap}$ and the high-temperature regime as $T\gg E_{gap}$. When \(T\ll\Delta_{\rm gap}\), thermal excitation of magnons is reduced due to their population being proportional to $(\exp(-E_{gap}/T))$, resulting in low-energy states dominating the dynamics. When $(T\sim E_{gap})$, a crossover occurs where thermally excited magnons begin to contribute significantly. For $(T\gg E_{gap})$, the gap no longer suppresses thermal excitations and the dynamics is dominated by a broad distribution of thermally populated magnons. On the other hand in the gapless phase (in the thermodynamic limit), any non-zero temperature can be considered as high temperature.

Fig.~\ref{fig3} shows the evolution of qubit coherence in the gapped ferromagnetic phase ($\Delta<-1$) and for the transition point $\Delta=-1$ (a). 
As reported in Ref.~\cite{Marcin_2025}, the phase transition at $\Delta=-1$ leads to the
gradual appearance of oscillations at zero temperature. At large $|\Delta|$
it is possible to show that the chain effectively acts as a two-level environment 
and the oscillations are the result of effectively two-qubit evolution 
which involves cyclic entangling and disentangling of the qubit and the chain~\cite{Marcin_2025}. At higher temperatures, the initial state of the chain contains admixtures of higher energy eigenstates of the chain Hamiltonian (\ref{xxz}) and, as expected leads to the damping of the oscillations. What is more surprising, however, is how gradually the increasing temperature induces damping. At finite, yet small temperatures compared the spin-chain spectral gap, $T \ll E_{gap}$, there is no discernible difference from the oscillations at zero temperature. Increasing the temperature leads to damping of these oscillations as can be seen in Fig.~\ref{fig3}. Interestingly, the oscillations persist even at infinite temperature. As $\Delta$ increases, the oscillation frequency increases, and the finite-temperature curves gradually converge toward the zero-temperature curve, as observed in the dip of the gapped phase (panel 2(d)). In the limit ($\Delta\rightarrow -\infty$), the spectral gap scales as ($E_{\mathrm{gap}}\sim J|\Delta|$) and therefore becomes infinitely large. As a result, thermal excitation is exponentially supressed at every finite temperature and excited states
remain essentially unoccupied. Increasing the temperature by any finite amount has negligible effect in the $\Delta\rightarrow -\infty$ limit and the qubit continues to oscillate coherently and the chain is effectively restricted to its two ground states.

Fig.~\ref{fig4} shows the decay of qubit coherence in the gapless phase of the spin chain for various negative $\Delta$ values, as well as $\Delta = 0$. Here the chain behaves like a large environment already at zero temperature, leading to dephasing which is well described by the Born approximation, but does display non-Markovian features~\cite{Marcin_2025}. In this phase, the effect of temperature is rapid and already low temperatures lead to decay which is quantitatively similar to the infinite-temperature curves. When the value of $\Delta$ is reasonably far from the transition point (panels (a), (b), and (c)), the higher-temperature curves at large times resemble exponential decay, suggesting that non-Markovian features at large times disappear with increasing temperature. The $\Delta = -0.9$ curves in panel (d) behave less predictably, yielding faster decoherence for some finite temperatures than for infinite temperature. This is due to the closeness of $\Delta$ value at the phase transition. Incidentally, the effects resulting from the finite length of the chain are visible in the low temperature curve in panel (c) at around $t = 80$ $1/J$, which suggest that results beyond this time do not reliably predict the behavior of an infinite chain. 

In the gapless phase, at low temperatures and weak qubit-chain coupling, the effect of temperature on the qubit decoherence can be rationalized as follows. Within the Born approximation, i.e.~for sufficiently weak coupling $g$, the time evolution of the probe qubit coherence is given by~\cite{Marcin_2025}
\begin{equation}
[\rho_{01}^{(2)}(t)]_{T} =
\rho_{01}(0)
\exp\left[
-\frac{g^{2}}{2}
\int_{0}^{t} d\tau
\int_{-\tau}^{\tau} dt'
{[S_{zz}(t)]}_T
\right],
\label{eq:rho}
\end{equation}
which expresses the qubit coherence dynamics in terms of the spin-chain correlation function ${[S_{zz}(t)]}_T=\big\langle S^{z}_{M}(t')S^{z}_{M}(0)\big\rangle_T$. At low-temperatures, where an effective low-energy description of the spin-chain in terms of Tomonaga–Luttinger liquid theory~\cite{Giamarchi_2003,Gogolin_1998,Haldane_1981} can be used, this correlation function is given by ~\cite{Giamarchi_2003,Cazalilla_2004} 
\begin{equation}
{[S_{zz}(t)]}_T \sim A \left[\frac{\pi T}{sinh(\pi T t)}\right]^2 + B \left[\frac{\pi T}{sinh(\pi T t)}\right]^{2K}.
\label{eq:szsz_temp}
\end{equation}
This depends on $\Delta$ through the Luttinger liquid parameter $K=\pi/[2(\pi-\cos^{-1} \Delta)]$~\cite{Giamarchi_2003}, which determines the strength of interactions between the elementary spin excitations and controls the asymptotic behavior of the correlation function. The constants $A$ and $B$ depend on $J$ and $\Delta$, and they control the relative magnitude of the corresponding (so-called smooth and staggered) contributions, respectively. We also note in passing that for $\Delta=0$, ${[S_{zz}(t)]}_T$ can be computed exactly using Jordan-Wigner transformation, which in the thermodynamic limit is given as 
\begin{equation}
    {[S_{zz}(t)]}_T=\left[{\displaystyle \int_{-J}^{+J}}d\omega\frac{f(\omega/T)}{\pi\sqrt{J^2-\omega^2}}e_{}^{i \omega t}\right]^2,
\end{equation}
with $f(x)=\left(e_{}^{x}+1\right)^{-1}$. For $T\ll J$, this gives ${[S_{zz}(t)]}_T\sim -\frac{1}{\pi^2}\left[\frac{\pi T}{sinh(\pi T t)}\right]^2$ consistent with the above Luttinger liquid field theoretic expression.

The behaviour of the qubit decoherence depends strongly on the sign and magnitude of $\Delta$. For the attractive case ($-1 < \Delta < 0$), $K > 1$ and for large times, $t\gg1/T$,  ${[S_{zz}(t)]}_T \sim A (2 \pi T)^{2} e^{-2\pi T t}$. Using this in Eq. (8), we get $[\rho_{01}^{(2)}(t)]_{T} \sim \exp\left[-\Gamma t \right]$ with the asymptotic decoherence rate given by $\Gamma = 2\pi A T g^{2}$. Hence, at sufficiently low temperatures and at weak qubit-chain coupling, qubit coherence decays exponentially at large times.

When comparing Figs \ref{fig3} and \ref{fig4} a qualitative difference in the 
qubit evolution is evident. In the gapless phase $-1<\Delta <0$ for large time ($t \gg 1/T$), the qubit decoherence follows an exponential decay. However, in the gapped ferromagnetic phase ($\Delta<-1$), the presence of a finite excitation gap qualitatively changes the dynamics, and the decoherence exhibits damped oscillatory behavior due to finiteness of $E_{gap}$. The effect of temperature has some unifying effect on the two types of curves, but this is much slower than expected. Although it is not a realistic expectation that the two types of curves would be distinguishable at very high temperatures (since the remnants of the oscillations in the gapped phase are very small at infinite temperatures), they should be visible at reasonably high temperatures, characteristic for experimental studies.

\section{\label{sec:summary}Conclusions}
In this work, we explored the potential of a single qubit probe to characterize the critical properties of a many body spin chain even when the chain is at finite temperature. Here the qubit is connected locally to the thermal spin chain instead of global access to the chain degrees of freedom. In particular, we examined how the decoherence dynamics of the qubit probe evolve as the control parameter of the chain is varied across the phase boundary. Our results reveal clear signatures of the underlying QPT encoded in the probe’s dynamics. 

In general QPTs are rigorously defined in the zero-temperature limit, where the ground-state properties undergo a qualitative change at the critical point. As the temperature increases, this conventional indicators of criticality are typically washed out due to thermal fluctuations. Nevertheless, we found that the distinct decoherence signatures exhibited by the qubit probe in the gapped and gapless phases of the chain remain remarkably robust against thermal effects.

We observed that the oscillations which are characteristic for the gapped phase at zero temperature persist when the temperature is increased. Moreover, the damping of the oscillations (which is accompanied by decrease in their frequency) occurs surprisingly slowly, and some remnants of the oscillations are still present at infinite temperatures. In comparison, the decoherence curves in the gapless phase behave according to expectations for the situation when the qubit evolves due to the interaction with a large environment. Here, the effect of temperature is stronger, and even at relatively low temperatures, the finite-temperature evolution closely resembles the infinite-temperature qubit evolution. We also observe distinct temperature dependences of the decoherence rate on the attractive and repulsive sides of the gapless phase. In the attractive regime, the decoherence rate increases linearly with $T$, whereas in the repulsive regime it has a nonlinear power-law dependence on temperature (see Appendix~\ref{sec:Appendix4} for details).

The proposed protocol could be implemented using NV centers in diamond~\cite{Casola_2018, Rovny_2024, rovny2024sensing}. The NV electronic spin could serve as the probe qubit, while the spin chain could be realized by a neighboring magnetic system or engineered spin arrays. NV based sensing has proven effective for probing thermally excited spin systems and many-body magnetic dynamics over a broad range of temperatures~\cite{Gaebel_2006, Toyli_2012,Brendan_2020,Finco_2021}. NV based experiments have also directly detected magnetic phase transitions through enhanced critical fluctuations~\cite{Wu_2025,Zhu_2025}. These results provide experimental support for using a single NV spin to probe the finite-temperature critical behavior of a nearby spin chain. This proposed protocol is also compatible with several quantum-simulation platforms. Ultra-cold atoms in optical lattices and trapped-ion systems allow the realization of tunable spin-chain Hamiltonians with controllable interaction strengths.~\cite{Trotzky_2008,Friedenauer_2008,Kim_2010,Simon_2011,Ma_2011,Islam_2011,Arrazola_2016,Monroe_2021}

In future this protocol can be extended to multiple probe qubits coupled to the spin chain~\cite{von_2020,Rovny_2022,Le_2025,Rovny2025,Hossein_2026}. Such a multi-probe configuration would enable the measurement of spatial correlations and collective signatures of criticality within the spin chain. By exploiting these additional degrees of freedom, it could enhance sensitivity and provide access to information beyond that available from a single local probe. Aditionally, decoherence dynamics of qubit probes in strongly coupled regimes can be studied to probe non-Gaussian fluctuations in many-body systems~\cite{Sung_2019,Norris_2016,Xia_2025,Curtis_2026}.
In that context, it is interesting to ask how far quantum critical information can survive as temperature increases, and whether a universal temperature scale exists beyond which local probes can no longer reliably identify the underlying QPT. Understanding this quantum to thermal crossover would provide further insight into the robustness of critical correlations in many-body systems.

\begin{acknowledgments}
S.D. acknowledges support from the European Union and the Czech Ministry
of Education, Youth and Sports (MEYS) (Project: MSCA Fellowship CZ FZU III, No. CZ.02.01.01/00/22010/0008598) and the computational resources provided by the e-INFRA CZ project (ID No. 90254), also supported by MEYS. M.T.M. and H.K.Y. are supported by the Royal Society University Research Fellowship \texttt{URF-T-261001}. S.D. thanks H.K.Y. and M.T.M. for their kind hospitality during his visit to King’s College London in the course of this work. 

\end{acknowledgments}

\appendix
\section{\label{sec:Appendix1}Proof that $\rho_{01}(t)^*=\rho_{01}(t)$ for $\epsilon=0$}

For $\epsilon=0$, the total Hamiltonian (\ref{total})
can be 
written in a block-diagonal form~\cite{Thomas_2006},
\begin{equation}
\label{atotal}
H=|0\rangle \langle 0| \otimes H_{+} + |1\rangle \langle 1| \otimes H_{-},
\end{equation}
where 
\begin{equation}
\label{acond}
H_{\pm}=H_{C}\pm \frac{g}{2}S_{M}^z.
\end{equation}
The structure of this Hamiltonian conveys that the
environment evolves under one of the two conditional Hamiltonians depending on the qubit state $|0\rangle$ or $|1\rangle$.

Using the conditional environmental Hamiltonians (\ref{acond}), the equation 
that governs the evolution of coherence (for $\epsilon=0$) is 
\begin{equation}
	\rho_{01}(t)=\frac{{\rm Tr}\left[e^{-i H_{-}t} e_{}^{-\beta H_{C}}e^{+i H_{+}t}\right]}{{\rm Tr}\left[e_{}^{-\beta H_{C}}\right]},
\end{equation}
so obviously
\begin{equation}
	\rho_{01}(t)^*=\frac{{\rm Tr}\left[e^{-i H_{+}t} e_{}^{-\beta H_{C}}e^{+i H_{-}t}\right]}{{\rm Tr}\left[e_{}^{-\beta H_{C}}\right]}.
\end{equation}

Using the spin-inversion symmetry operator, $\mathbb{U}=\bigotimes_{l=1}^{L} 2 S_l^x$, it is easy to see that $\mathbb{U} H_{C}\mathbb{U}^\dagger=H_{C}$ and  $\mathbb{U} H_{\pm}^z\mathbb{U}^\dagger = H_{\mp}$ for $\epsilon=0$. Using this, along with the cyclic invariance of the trace, we get
\begin{equation}
	\rho_{01}(t)^*=\frac{{\rm Tr}\left[e^{-i H_{-}t} e_{}^{-\beta H_{C}}e^{+i H_{+}t}\right]}{{\rm Tr}\left[e_{}^{-\beta H_{C}}\right]}
    =\rho_{01}(t).
\end{equation}
\begin{figure}[b]
\begin{center}
\includegraphics[width=3.5 in]{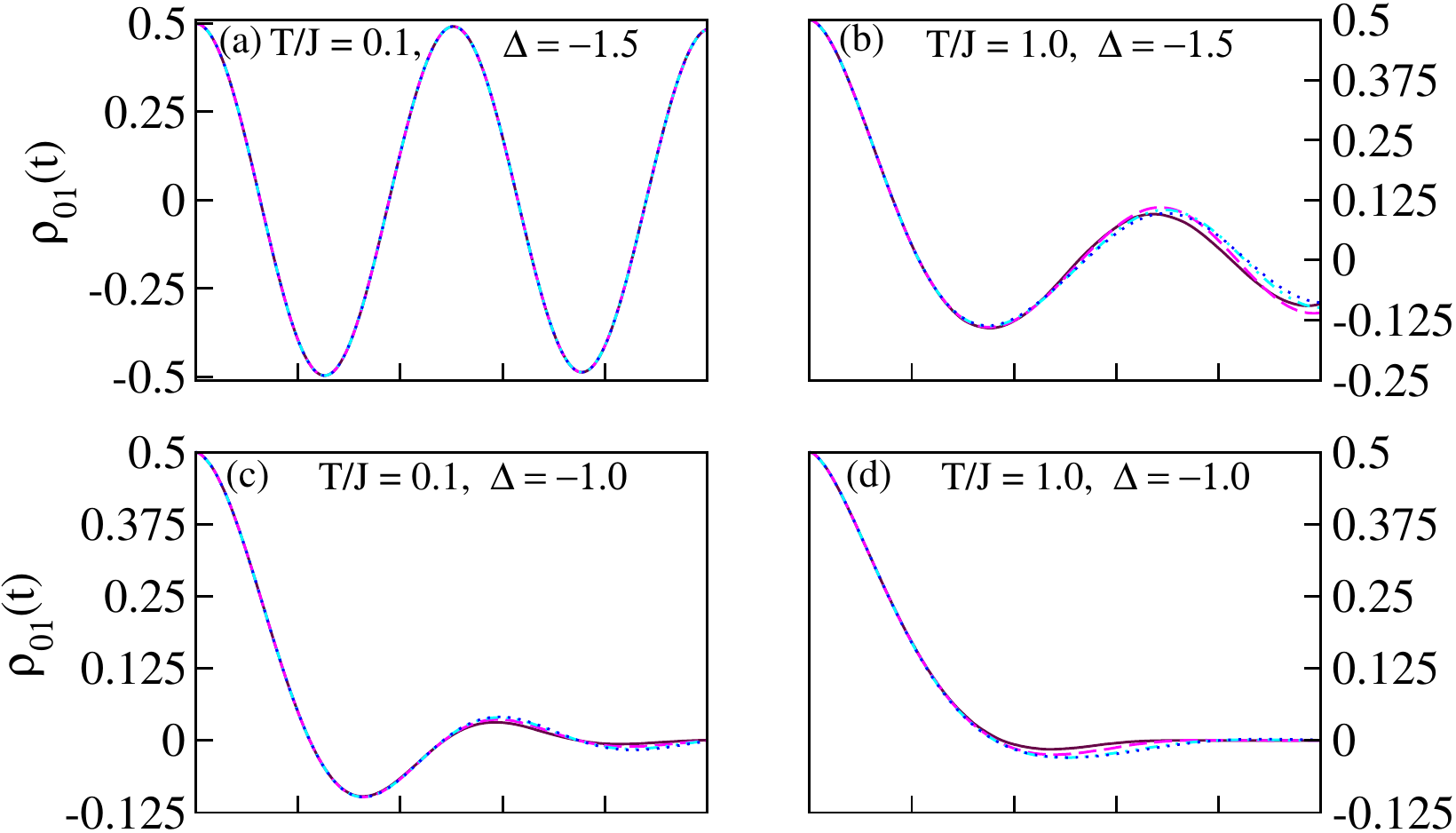}
\par\vspace{0.3cm}
\includegraphics[width=3.4 in]{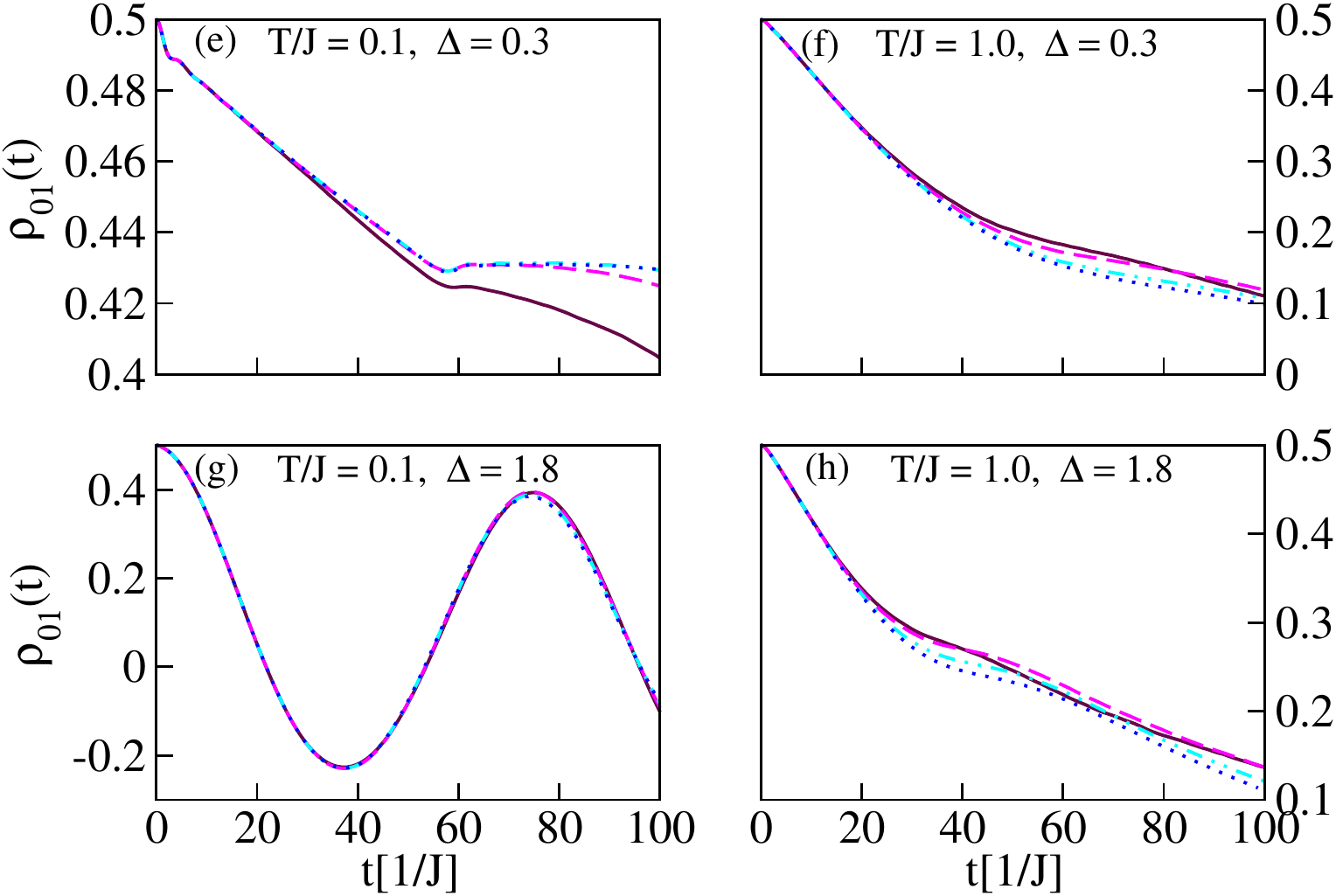}
    \caption{\label{fig:plt34} Convergence of $\rho_{01}(t)$ with respect to bond dimension $\chi$ for different $T/J$  and $\Delta$ values for a system size of $L = 64$. Different curves correspond to different bond dimension values: solid maroon lines for $\chi = 128$, dashed magenta lines for $\chi = 256$,
    dotted-dashed cyan lines for $\chi = 512$ and dotted blue lines for $\chi = 768$. }
    \label{fig5}
\end{center}
\end{figure}
\begin{figure*}[t]
\begin{center}
\includegraphics[width=6.4 in]{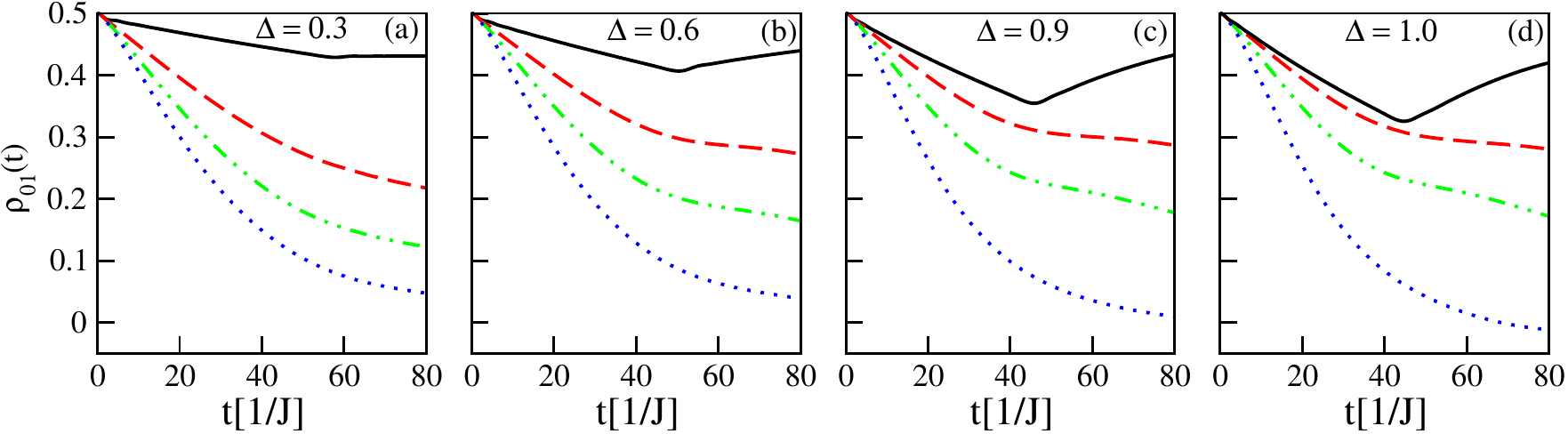}
\par\vspace{0.3cm}
\includegraphics[width=6.7 in]{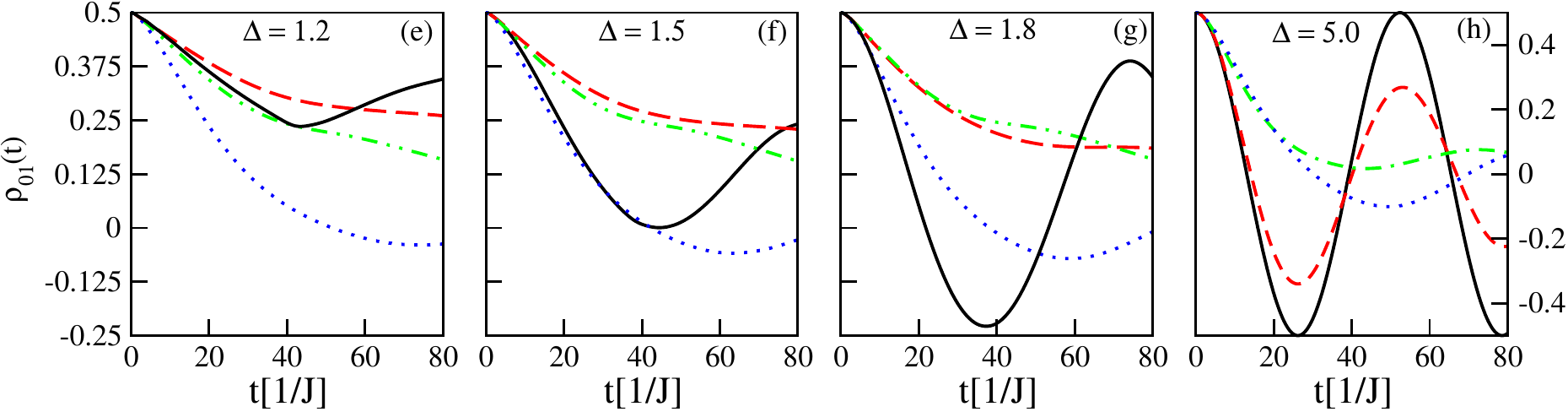}
    \caption{ Qubit decohernce $\rho_{01}(t)$ for a system size of $L = 64$ and bond dimension $\chi=768$ for different $T/J$ and $\Delta$ values. Different curves correspond to different temperature values: solid black lines for $T/J = 0.1$, dashed red lines for $T/J = 0.5$, dotted-dashed green lines for $T/J = 1.0$ and dotted blue lines for $T/J = \infty$. All panels show $\rho_{01}(t)$ as a function of time $t$. Figures (b), (c) and (d) are plotted using the same scales as figure (a). Figures (f) and (g) are plotted using the same scales as figure (e), whereas figure (h) is plotted using a different vertical scale.}
    \label{fig6}
\end{center}
\end{figure*}

\section{\label{sec:Appendix2}Purification Method}
In the purification method, an ancilla is introduced as a complement to each physical site of the spin chain. Each ancilla has the same Hilbert-space dimension as its corresponding physical spin. Therefore, for a spin chain of length $L$, we rewrite Eq.(\ref{eq:lecho}) as the overlap of two pure states defined on an extended chain of length $2L$, with $L$ sites for the physical spins and $L$ for identical ancillas~\cite{Verstraete_2004, Czarnik_2012, Czarnik_2019, Suzuki_1985,Feiguin_2005,Zwolak_2004} as follows, 
\begin{eqnarray}
\rho_{01}(t) &=&\frac{e_{}^{i \epsilon t}}{2}\frac{{\rm Tr}\left[e_{}^{i H_+ t}e_{}^{-i H_- t}e_{}^{-\beta H_C}\right]}{{\rm Tr}\left[e_{}^{-\beta H_C}\right]}\nonumber\\
&=&\frac{e_{}^{i\epsilon t}}{2}\langle \Psi_+(\beta/2,t)\mid \Psi_-(\beta/2,t)\rangle,
\end{eqnarray}
where
\begin{eqnarray}
\mid \Psi_\pm(\beta,t)\rangle &=&\frac{\mid\tilde{\Psi}_\pm(\beta,t)\rangle}{\|\tilde{\Psi}_\pm(\beta,t)\rangle\|}\nonumber
\end{eqnarray}
with
\begin{eqnarray}
\mid \tilde{\Psi}_-(\beta,t)\rangle &=&e_{}^{-i t \left(H_-\otimes \mathbb{I}-\mathbb{I}\otimes H_-\right)}e_{}^{-\frac{\beta}{2}\left(H_C \otimes \mathbb{I}+\mathbb{I}\otimes H_C\right)} \mid \mathcal{I}\rangle\nonumber\\
\mid \tilde{\Psi}_+(\beta,t)\rangle &=&e_{}^{-i t \left(H_+\otimes \mathbb{I}-\mathbb{I}\otimes H_-\right)}e_{}^{-\frac{\beta}{2}\left(H_C \otimes \mathbb{I}+\mathbb{I}\otimes H_C\right)} \mid \mathcal{I}\rangle\nonumber\\
\end{eqnarray}
where the purified infinite-temperature state $\mid \mathcal{I}\rangle$ is given as
\begin{equation}
\begin{aligned}
\mid \mathcal{I}\rangle
&= \frac{1}{2^{L/2}}
\sum_{\sigma_1=\uparrow,\downarrow}\cdots
\sum_{\sigma_L=\uparrow,\downarrow}
(-1)^{\sum_k \delta_{\sigma_k \downarrow}}
\\[-2pt]
&\qquad
\mid \sigma_1,\cdots,\sigma_L\rangle_{\text{physical}}
\mid \bar{\sigma}_1,\cdots,\bar{\sigma}_L\rangle_{\text{ancilla}}
\\[-2pt]
&\equiv
\otimes_{k=1}^{L}
\left[
\frac{
\mid \uparrow\rangle_{k,\text{physical}}
\mid \downarrow\rangle_{k,\text{ancilla}}
-
\mid \downarrow\rangle_{k,\text{physical}}
\mid \uparrow\rangle_{k,\text{ancilla}}
}{\sqrt{2}}
\right].
\end{aligned}
\end{equation}
with $\bar{\sigma}=\downarrow/\uparrow$ if $\sigma=\uparrow/\downarrow$.
The real and imaginary time evolutions to obtain $|\psi_\pm(\beta,t)\rangle$ from $|\mathcal{I}\rangle$ are carried out by employing Time Dependent Variational Principle (TDVP)~\cite{Haegeman_2011,Haegeman_2016,Mingru_2020} and approximately representing intermediate states during evolution, $|\Psi_\pm(\beta,t)\rangle$, as Matrix Product States (MPS)~\cite{Schollwock_2011,Orus_2014} with bond-dimensions not exceeding a suitably chosen value $\chi$. The evolution is carried out in small time step, $dt$.

\section{\label{sec:Appendix3}Convergence Test}
Numerical exactness of the dynamics generated by TDVP-MPS is obtained with respect to the bond dimension $\chi$. In Fig.~\ref{fig5}, we present a comparison of the calculated $\rho_{01}(t)$ using bond dimensions ranging from $\chi=128$ to $\chi=768$ for various values of $T/J$ and $\Delta$. The results are shown for a system of size $L=64$ with open boundary conditions (OBC).
We retain the truncation error to less than  $10^{-8}$.  We observe that as the bond dimension increases, results of $\rho_{01}(t)$ show clear convergences. We limit the bond dimension to a maximum of 768 for our calculations as reported in the main text. The numerical results in this work were obtained using the open-source library ITensor~\cite{itensor}. 

\section{\label{sec:Appendix4} Repulsive regime ($\Delta > 0$)}
Fig.~\ref{fig6} shows the dynamics of qubit coherence ($\rho_{01}(t)$) for the spin-chain size, $L=64$ in the repulsive regime ($\Delta >0$). In the gapless Luttinger-liquid (LL) phase, Figs. (a)–(d) show that the qubit probe decoheres monotonically with time except at $T=0.1/J$, where recoherence happens at times that depend on $\Delta$. This recoherence phenomenon can be attributed to the finite length of the spin-chain~\cite{Marcin_2025}. At temperatures larges than $T=0.1/J$, this recoherence does not appear. Before this recoherence time, the the absolute value of the qubit coherence at any fixed time decreases as temperature increases.

More insights into the dependence of the qubit decoherence at low temperatures in the repulsive gapless regime can be gained using the low energy Luttinger liquid theory, similar to that of the attractive regime in the main text. For anisotropy values $0 < \Delta < 1$, the Luttinger parameter satisfies $K < 1$. Using this in the Eq. (\ref{eq:szsz_temp}), for large times, $t\gg1/T$,  ${[S_{zz}(t)]}_T \sim B (2\pi T)^{2K} e^{-2\pi K T t}$. Using this in Eq. (\ref{eq:rho}), we get $[\rho_{01}^{(2)}(t)]_{T} \sim \exp\left[-\Gamma t \right]$ with the asymptotic decoherence rate given by $\Gamma = [g^{2} B {(2\pi T)^{2K-1}}]$. Hence, at sufficiently low temepratures and at weak qubit-chain coupling, qubit coherence decays exponentially at large times. Noticeably, the decoherence rate is non-linear in temperature, in contrast to the attractive regime.

Figs. (e)-(h) show dynamics of qubit coherence in the repulsive anti-ferromagnetic gapped regime ($\Delta>1$). At $\Delta = 1$ corresponds to isotropic antiferromagnetic Heisenmebrg model, which marks  a Berezinskii-Kosterlitz-Thouless (BKT) quantum critical point separating the gapless Luttinger-liquid phase from the gapped antiferromagnetic phase. For $\Delta=1^{+}$, i.e., close to the transition point, a gap opens with an exponentially weak dependence on the distance from criticality, of the form
$E_{gap} \sim 4\pi J*\mathrm{exp}\left[\frac{-\pi^{2}}{2\sqrt{2(\Delta -1)}}\right]$~\cite{Cloizeaux_1966}. Consequently, at finite temperature, the gap remains much smaller than the thermal energy over a broad range of $\Delta$ near $\Delta=1$. As a result, the qubit decoherence exhibits a considerably less pronounced change around $\Delta=1$ compared with the behavior near $\Delta=-1$. Similar to case of attractive regime in main text, as ($\Delta \rightarrow \infty$), the spectral gap increases proportionally to $J|\Delta|$, leading to a strong suppression of thermal excitations. As a result, the qubit maintains its coherent oscillations even at finite temperatures, and the finite-temperature dynamics progressively converge to the zero-temperature behavior.

\bibliography{bibliography}
\end{document}